\documentclass[aps,pra,reprint,twocolumn,superscriptaddress]{revtex4-1}
\usepackage{appendix}
\usepackage{epsfig}
\usepackage{graphicx}
\usepackage{amsfonts}
\usepackage{amsmath}
\usepackage{psfrag}
\usepackage{amssymb,amsmath,amsthm}
\usepackage{verbatim}
\usepackage{soul}
\usepackage[normalem]{ulem}
\usepackage{color}
\usepackage{float}

\begin{document}

\title{Radiation signal accompanying the Schwinger effect}
\author{I.~A.~Aleksandrov}
\affiliation{Department of Physics, Saint Petersburg State University, 199034 Saint Peterburg, Russia}
\affiliation{Ioffe Institute, 194021 Saint Peterburg, Russia}
\author{A.~D.~Panferov}
\affiliation{Saratov State University, 410026 Saratov, Russia}
\author{S.~A.~Smolyansky}
\affiliation{Saratov State University, 410026 Saratov, Russia}
\affiliation{Tomsk State University, 634050 Tomsk, Russia}
 

\begin{abstract}
The properties of the forced oscillations of electron-positron plasma (EPP) generated from vacuum under the action of a short laser pulse are considered. Calculating the density of the conduction and polarization currents within the quantum kinetic approach, we demonstrate the presence of plasma oscillations at the frequency of the external field and its odd harmonics. It is expected that radiation generated by these plasma oscillations can be observed outside the interaction region, for example, outside the focal spot of two counterpropagating laser beams, and can serve as an indicator of the Schwinger mechanism of the EPP creation from vacuum.
\end{abstract}

\maketitle

\vskip 1cm

\section{Introduction}

Viewing vacuum not as a mere scene where material objects interact with each other was a necessary step towards a self-consistent quantum field theory.
Nontrivial properties of the quantum vacuum manifest themselves, for instance, in the phenomenon of Schwinger pair production~\cite{Schwinger:1951nm, sauter_1931, euler_heisenberg}.
To observe this effect, one has to approach the regime of extremely large electric field strength $E_\text{c} = m^2 c^3 /(|e|\hbar) \approx 1.3 \times 10^{16}~\text{V/cm}$ ($m$ and $e<0$ are the electron mass and charge, respectively).
The possibility of reaching this value, albeit for a very short time, appears due to current developments of the technology for generating ultrahigh-power laser pulses.
To achieve the critical value $E_\text{c}$, it is necessary to ensure the energy flux density in the focal spot of the laser setup close to $2.3 \times 10^{29}~\text{W/cm}^2$.
Presently, intensity of about $10^{23}~\text{W/cm}^2$ ($E \approx 0.001E_\text{c}$) is considered to be an attainable regime within a number of projects under construction and commissioning~\cite{Zou:2015, Weber:2018, Zamfir:2014, Zeng:2017, Sung:2017}.
For the next-generation projects \cite{Meyerhofer:2014, Shen:2018, Kawanaka:2016, Mourou:2011, Bashinov:2014}, the expected values of this parameter are in the range $10^{24}$--$10^{25}~\text{W/cm}^2$ ($E \approx 0.01E_\text{c}$).
Under these conditions, the question of setting up experiments takes on practical significance.

Usually, direct detection of positrons or detection of characteristic signatures of positron annihilation is considered as a way of observing the Schwinger effect. Now we possess a detailed information on the dependence of the characteristics of primary pairs on the parameters of the active field. Basically, it can be argued that the possible directions of the emission of the primary particles and their energies can be predicted with high accuracy and resolution. This should prevent an unambiguous interpretation of the recorded events. However, there will also be a masking background arising due to cascade processes (see, e.g., Refs.~{\cite{bell_kirk_2008, Fedotov:2010, elkina_2011, Nerush:2011, narozhny_fedotov_2015, Grismayer:2017}}) involving the primary particles themselves or inevitable impurities.

In this study, we use the quantum kinetic approach and show that the polarization of the vacuum state in the presence of a strong alternating electric field can manifest itself in the form of re-emission of part of its energy both at its fundamental frequency and at its odd high-frequency harmonics. In particular, we will demonstrate that subcritical vacuum polarization results in the third harmonic in the radiation spectrum. 

This study complements recent investigations of the radiation process described by the tadpole QED diagram~\cite{Karbstein:2015_2, Alex:2019, Gies:2018, Blinne:2019, Karbstein:2019, King:2018} (see also Refs.~\cite{DiPiazza:2005, Fedotov:PLA:2007, Narozhny:2007}). Our findings represent a next step towards understanding the polarization properties of quantum vacuum, which also give rise, for instance, to a remarkable phenomenon of vacuum birefringence (see, e.g., Refs.~\cite{Karbstein:2015, Karbstein:2016, Karbstein:2018} and references therein).

The structure of this paper is the following. In Sec.~\ref{sect:ke} we briefly describe a theoretical framework based on the quantum kinetic equations. In Sect.~\ref{sect:ldl} we discuss the temporal evolution of the conduction and polarization currents. Our main findings concerning quasiclassical radiation are presented in Sec.~\ref{sect:br}. In Sec.~\ref{sect:co} we provide a discussion. Throughout the text, we assume $c=\hbar=1$.

\section{Kinetic equation \label{sect:ke}}

For the description of the process of vacuum EPP production at $t\to\infty$ and also of the intermediate quasiparticle excitations and polarization effects in a strong electric field, we will employ the kinetic equation (KE) for the case of a uniform linearly polarized field $E(t)=-\dot{A}(t)$ with a vector potential $A^\mu(t)=(0,0,0,A(t))$~\cite{Bialynicki-Birula:1991, Grib:1994, Schmidt:1998, kluger_1998} (for more details, see also Refs.~\cite{Smol:2016,qke:2020}):
\begin{equation}\label{ke}
\dot f(\mathbf{p} ,t) = \frac{1}{2} \lambda(\mathbf{p}
,t )\int\limits^t_{t_\text{in}} dt^{\prime} \lambda(\mathbf{p} ,t^{\prime})[1-2f(\mathbf{p}
,t^{\prime})]\cos\theta(t,t^{\prime}),
\end{equation}
where $f(\mathbf{p} ,t)$ is the quasiparticle distribution function,
\begin{eqnarray}
\lambda(\mathbf{p},t) &=& \frac{e E(t)\varepsilon_{\bot}}{\varepsilon^{2}(\mathbf{p},t)} , \label{lambda}\\
\theta(t,t') &=& 2 \int \limits^t_{t'} d\tau \, \varepsilon (\mathbf{p} ,\tau). \label{phase}
 \end{eqnarray}
Here $\lambda$ is the amplitude of the vacuum transitions, and $\theta$ is a high-frequency phase describing the vacuum oscillations which are modulated by the external field.
The quasienergy $\varepsilon(\mathbf{p} ,t) = \sqrt{\varepsilon^2_{\bot}+ P^2}$ is defined by means of transverse energy $\varepsilon_\bot = \sqrt{m^2 + p^2_\bot}$ and longitudinal kinetic quasimomentum $P = p_\parallel -eA(t)$.
Here $p_\bot=|\mathbf{p_\bot}|$ is the magnitude of the vector $\mathbf{p_\bot}$ perpendicular to the field direction, and  $p_\parallel=p_3$ is the momentum component parallel to the external field.

We consider Eq.~(\ref{ke}) with zero initial conditions, i.e., $f(\mathbf{p} ,t_\text{in}) = 0$, and set $E(t_\text{in})=0$ and $A(t_\text{in})=A_\text{in}$.
It is also assumed that the electric field is switched off in the {\it out} state [$E(t\to\infty)=0$ and $A(t\to\infty) = A_{\rm out}$ can differ from $A_{\rm in}$].

The non-Markovian integro-differential equation (\ref{ke}) is equivalent to a system of three ordinary differential equations:
\begin{eqnarray}\label{ode}
 \dot{f} = \frac{1}{2}\lambda u, \quad \dot{u} = \lambda (1-2f) - 2 \varepsilon v, \quad \dot{v}= 2 \varepsilon u ,
\end{eqnarray}
where $u(\mathbf{p},t)$ and $v(\mathbf{p},t)$ are auxiliary functions describing vacuum polarization effects. The equations~(\ref{ode}) are to be considered with the initial conditions $f(\mathbf{p}, t_\text{in}) = u(\mathbf{p}, t_\text{in}) = v(\mathbf{p}, t_\text{in}) = 0$. This system is convenient for a numerical investigation of the problem.

The density of the quasiparticles, i.e., the number of the quasiparticles per unit volume, reads
\begin{equation}
\label{dens}
    n(t) = 2 \int [dp] f(\mathbf{p} ,t),
\end{equation}
where the factor 2 corresponds to spin degeneracy, and $[dp] \equiv d^3p/(2\pi)^3$. The total
current density can be decomposed into the sum of the conduction and polarization components~\cite{Bloch:1999},
\begin{equation}\label{cur_s}
j(t) = j^\text{cond}(t) + j^\text{pol}(t),
\end{equation}
where
\begin{eqnarray}\label{j_cond}
j^\text{cond}(t) &=& 2e\int [dp]\frac {P} {\varepsilon(\mathbf{p},t)} f(\mathbf{p},t) ,\\
j^\text{pol}(t) &=& -e\int [dp]\frac {\varepsilon_{\bot}} {\varepsilon(\mathbf{p},t)} u(\mathbf{p},t) .
\label{j_pol}
\end{eqnarray}
According to Eq.~(\ref{j_pol}), the function $u(\mathbf{p},t)$ can be interpreted as a current polarization function.
The function $v(\mathbf{p},t)$ determines the energy density of the vacuum polarization~\cite{Smol:2016, qke:2020}. Both of these components originate from single-particle anomalous correlation functions.

Let us discuss the convergence of the integrals in Eqs.~\eqref{j_cond} and \eqref{j_pol}. 
Assuming that $p \gg m$, i.e., $\varepsilon \gg m$, one can demonstrate that the asymptotic behavior of the functions $f(\mathbf{p}, t)$ and $u(\mathbf{p}, t)$ reads~\cite{Mamayev:1979} 
\begin{eqnarray} 
f \simeq f_4 &=& \frac{1}{16} \bigg ( \frac{\lambda}{\varepsilon} \bigg )^2, \label{eq:f_uv} \\ 
u \simeq u_3 &=& \frac{1}{4 \varepsilon}  \, \frac{d}{dt} \bigg ( \frac{\lambda}{\varepsilon} \bigg ), 
\label{eq:u_uv} 
\end{eqnarray} 
where the subscripts indicate the order of the corresponding terms within the power expansion with respect to $m/\varepsilon \ll 1$. From this, it follows that the conduction current~\eqref{j_cond} is already well defined, whereas the integral in Eq.~\eqref{j_pol} possesses a logarithmic divergence in the ultraviolet domain and should be regularized according to the subtraction procedure $u(t) \to u(t) - u_3 (t)$~\cite{Mamayev:1979}. 
In what follows, we will always take into account this counterterm providing finite values of the polarization current which will be denoted by $j^\text{pol}_\text{R} (t)$.

According to the KEs~(\ref{ke}) and (\ref{ode}), the distribution function $f(\mathbf{p},t)$ and the vacuum polarization functions  $u(\mathbf{p},t)$ and $v(\mathbf{p},t)$ incorporate temporal oscillations with the characteristic frequency of the external field (low frequency) and with the doubled frequency of the vacuum oscillations $2\sqrt{m^2 +
\mathbf{p}^2}$ (high frequency) depending on the momentum $\mathbf{p}$. Integrating the macroscopic densities~(\ref{dens}), (\ref{j_cond}), and (\ref{j_pol}) over momentum space flattens the high-frequency oscillations making the results slowly varying with $t$ and more representative.

Finally, we note that within a period of the external field action, one can represent the total current~(\ref{cur_s}) as a sum of the conduction and polarization components only formally as these cannot be distinguished experimentally. Basically, the vacuum polarization effects dominate in the densities mentioned above. In what follows, we will examine these effects by analyzing the currents (\ref{j_cond}) and (\ref{j_pol}).

\section{Particle density and currents\label{sect:ldl}}

\begin{figure*}[t]
\includegraphics[height=0.33\textwidth]{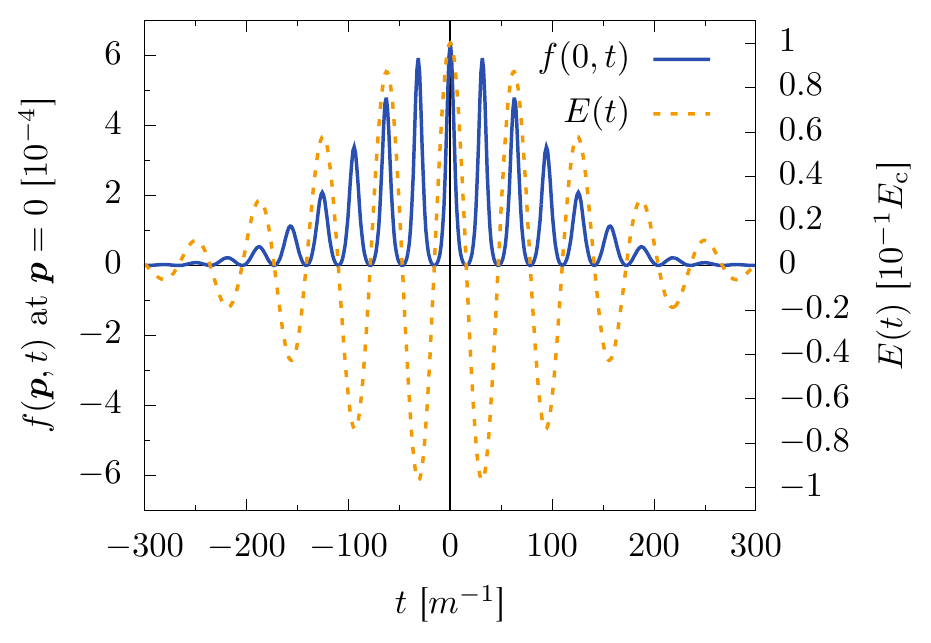}  \hfill
\includegraphics[height=0.335\textwidth]{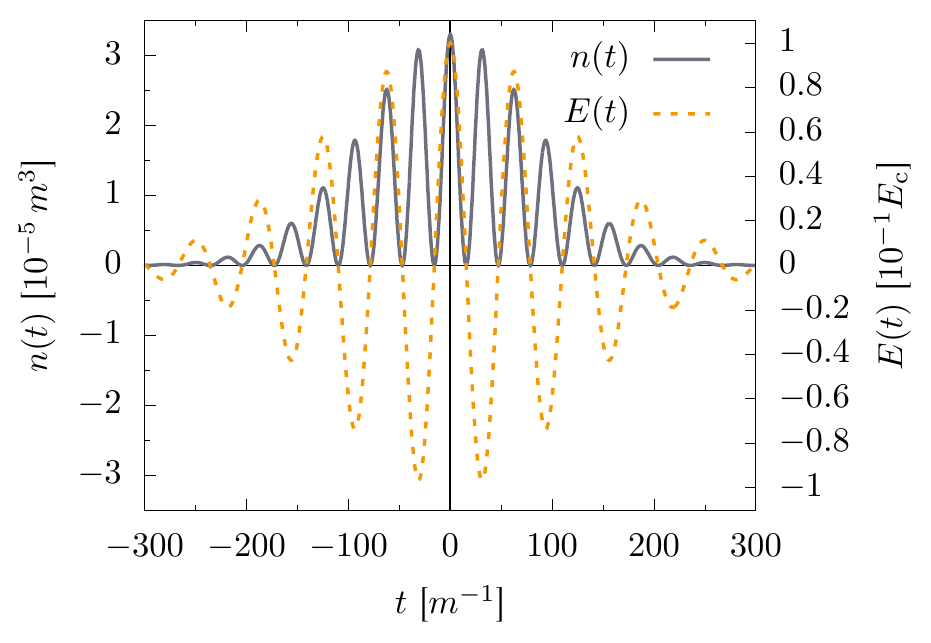} 
\caption{Distribution function $f(\mathbf{p},t)$ at $\mathbf{p}=0$ (left) and pair density (\ref{dens}) (right) as functions of time together with the external field strength $E(t)$. The field parameters are $E_0 = 0.1E_\text{c}$, $\Omega = 0.1 m$, $\sigma = 12$.
\label{fig:1}}
\end{figure*}

We consider the external electric background in the following form:
\begin{equation} \label{field} 
E(t) = E_0  \cos{ \Omega t }\ \mathrm{e}^{-t^2/(2\tau^2) }.
\end{equation}
We introduce also parameter $\sigma= \Omega \tau$ which is a dimensionless measure of the characteristic duration $\tau$ of the pulse governing the number of the carrier cycles. In this case, $t_\text{in} \to - \infty$.
The expression (\ref{field}) mimics the focal spot of two counterpropagating high-intensity laser pulses.

\begin{figure*}[t]
\includegraphics[height=0.33\textwidth]{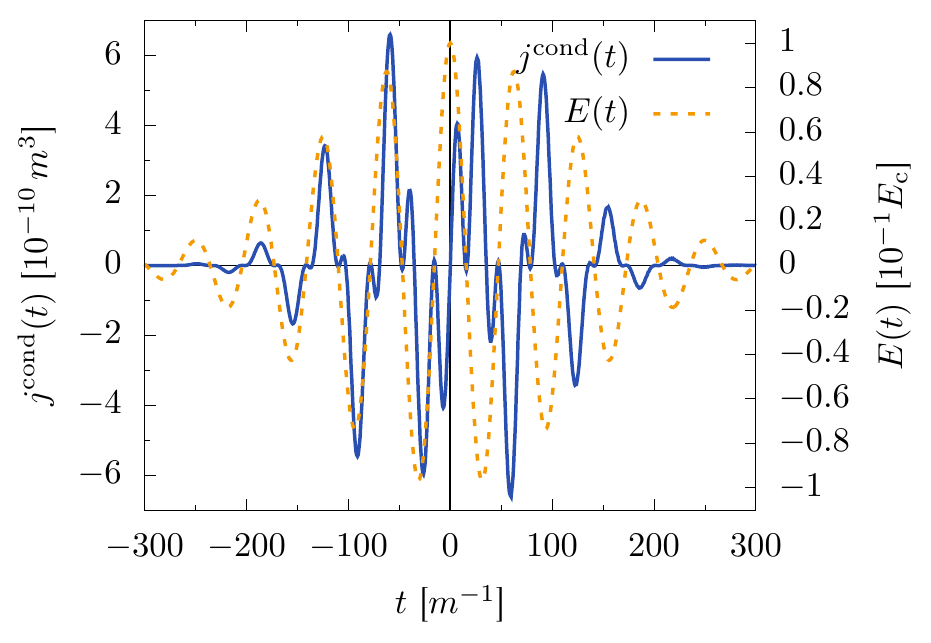} \hfill
\includegraphics[height=0.33\textwidth]{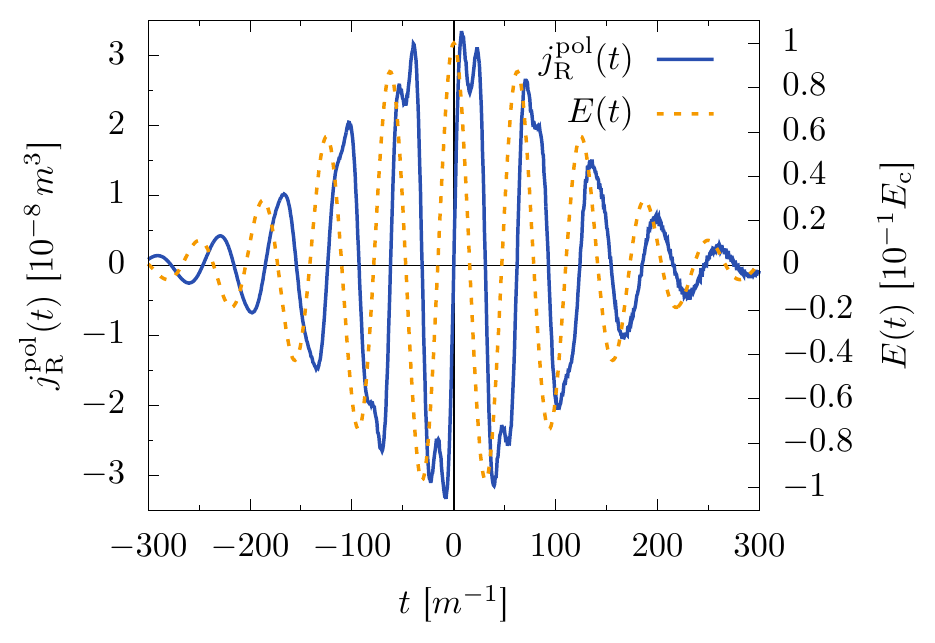}
\caption{Temporal dependence of the conduction current (left) and the polarization current (right) together with the external field strength $E(t)$. The field parameters are $E_0 = 0.1E_\text{c}$, $\Omega = 0.1 m$, $\sigma = 12$.
\label{fig:2}}
\end{figure*}

Our goal is to investigate the temporal behavior of the number density of quasiparticles~(\ref{dens}) and current densities (\ref{j_cond})--(\ref{j_pol}) during the action of the external electric pulse~(\ref{field}).
Unfortunately, for optical and near infrared frequencies this task is beyond the present computational capabilities, so we will work within the hard X-ray domain. We expect that the main findings of this study will also provide correct qualitative predictions in the region of lower frequencies. In our simulations, the external field frequency corresponds to the photon energy of $51$~keV ($\Omega = 0.1 m$).

The problem is solved in two stages. At the first stage, we calculate the distribution $f(\mathbf{p},t)$ according to the KE~(\ref{ke}) and then, at the second stage, we evaluate the integrals~(\ref{dens}), (\ref{j_cond}), and (\ref{j_pol}). The first stage is the most laborious since the function $f(\mathbf{p},t)$ is defined in four-dimensional space. By solving the system of equations~(\ref{ode}), one can obtain its temporal dependence only for a given set of parameters $\{p_1, p_2, p_3\}$. However, the axial symmetry reduces the dimension of the problem, so we assume $p_2 = 0$ leaving $p_1$ and $p_3$ as the transverse and longitudinal projections of the particle momentum, respectively ($p_1 = p_\perp$, $p_3 = p_\parallel$).

Our calculations were carried out for $E_0 = 0.1E_\text{c}$. Although the energy of the external-field quanta amounts only to $51$~keV, with such a high intensity, it turns out that the virtual electrons or positrons can gain about ten times more energy. Accordingly, the functions $f(\mathbf{p} ,t)$, $u(\mathbf{p} ,t)$,  and $v(\mathbf{p} ,t)$ were computed within the domain $0 \leq p_1 \leq p_\text{cut}$, $-p_\text{cut} \leq p_3 \leq p_\text{cut}$, where $p_\text{cut}=10.0m$. At the second stage, the integral characteristics, i.e., the number density~(\ref{dens}) and currents~(\ref{j_cond}) and (\ref{j_pol}), were calculated at each step of the temporal grid containing 1000 points within the laser pulse support, which provided a high temporal resolution.

\begin{figure*}[t]
\includegraphics[height=0.37\textwidth]{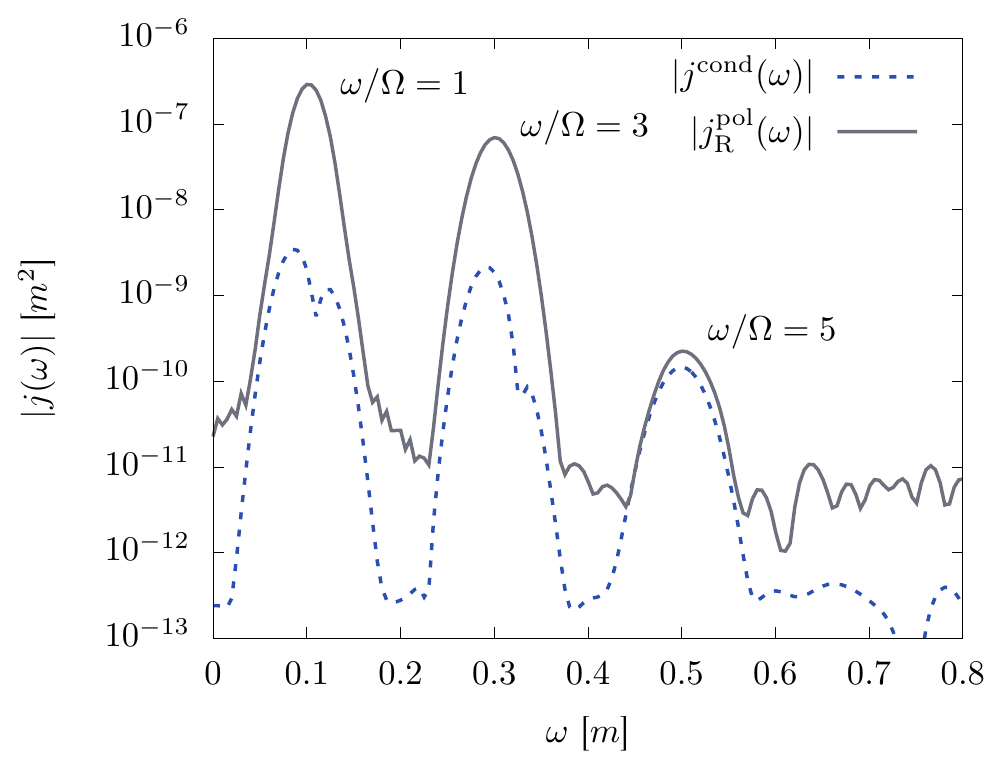}~~~~~\includegraphics[height=0.36\textwidth]{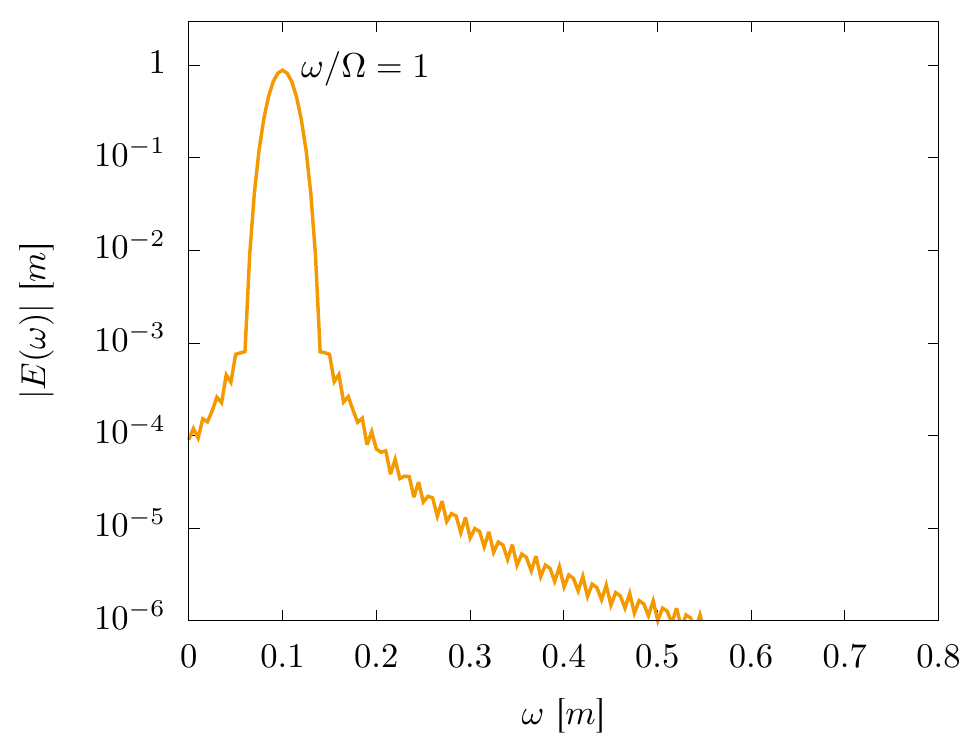}
\caption{Spectrum of the conduction and polarization currents (left) and that of the external field $E(t)$ (right). The field parameters are $E_0 = 0.1E_\text{c}$, $\Omega = 0.1 m$, $\sigma = 12$. \label{fig:3}}
\end{figure*}

Figure~\ref{fig:1} allows one to compare the temporal evolution of the distribution function at $\mathbf{p}=0$ 
with the behavior of the pair number density~(\ref{dens}). 
In these graphs we also depict the temporal dependence of the field strength~(\ref{field}) in order to demonstrate that the particle density evolution is qualitatively similar to $E^2(t)$~\cite{Kravtcov:2017}. The high-frequency oscillations become visible when one employs a logarithmic scale.

The temporal dependence of the conduction and polarization currents is displayed in Fig.~\ref{fig:2}. The results for the conduction current were obtained using the definition (\ref{j_cond}). The polarization current was calculated in accordance with Eq.~(\ref{j_pol}) taking into account the counterterm (\ref{eq:u_uv}) for regularization. First, one observes that the polarization current is about two orders of magnitude larger than the conduction one.
Second, the conduction current exhibits a much more complicated temporal behavior containing double spikes of alternating direction [Fig.~\ref{fig:2}~(left)]. 
The evolution of the polarization current is simpler [see Fig.~\ref{fig:2}~(right)]: its temporal dependence $j^\text{pol}_\text{R}(t)$ approximately reproduces a rescaled plot of $E(t)$ [Eq.~(\ref{field})] shifted by $\Delta t = \pi/(2\Omega)$.

The main part of the present study is devoted to the analysis of the spectral content of the calculated currents. 
In order to investigate this issue and perform the corresponding calculations, we utilized the standard routines \texttt{Periodogram} of Mathematica~\cite{mathematica}.
For better visibility of the results, we averaged the data from the partitions of the original list and smoothed the values with a window function of the Hann type. The resulting absolute values are presented in Fig.~\ref{fig:3}~(left). The both curves clearly demonstrate the presence of higher odd harmonics ($\omega/\Omega = 1$, $3$, $5$, ...).
The third harmonic of the polarization current is well pronounced and is about two orders of magnitude larger than that of the conduction current.
For the fifth harmonic, the amplitudes of the currents turn out to be very close in magnitude. For comparison, in Fig.~\ref{fig:3}~(right) we depict analogous Fourier transform of the external field (\ref{field}) in which there are obviously no additional harmonics.
This emphasizes that the appearance of high-frequency harmonics in the generated currents is a result of a complex nonlinear response of the physical vacuum.

\section{Backreaction problem and radiation\label{sect:br}}

The EPP currents generate the internal electric field $E_\text{int}(t)$ which obeys Maxwell's equation
\begin{equation}
\label{Max}
\dot E_\text{int}(t) = -j(t) = -j^\text{cond}(t) - j^\text{pol}_\text{R}(t).
\end{equation}
\begin{figure*}[t]
\includegraphics[height=0.36\textwidth]{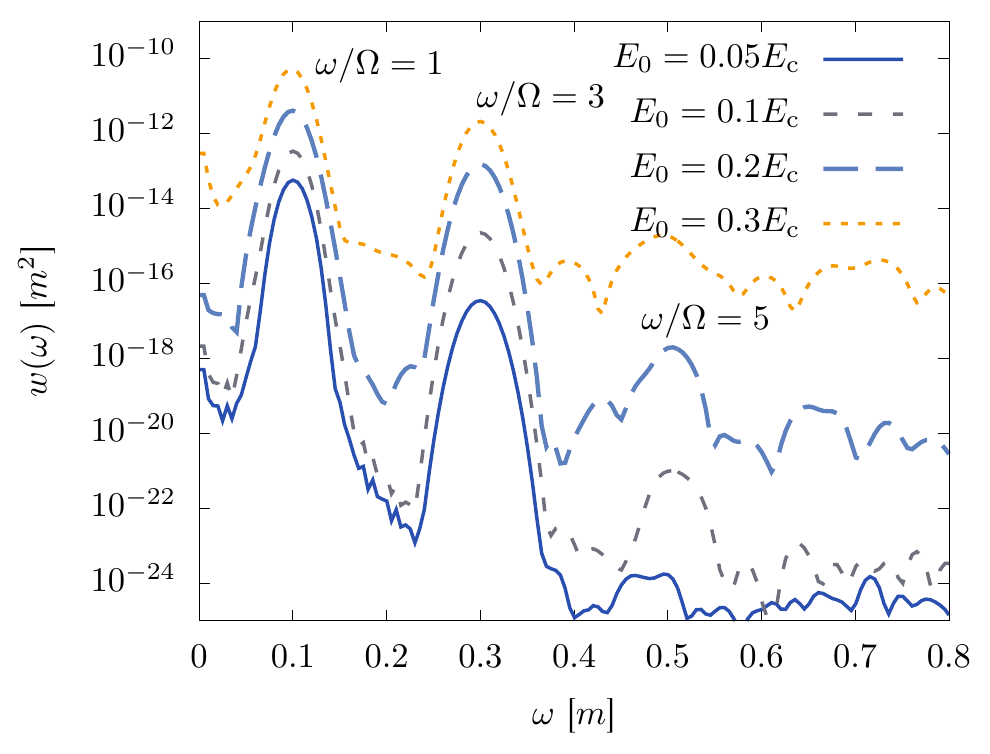}~~~~~\includegraphics[height=0.37\textwidth]{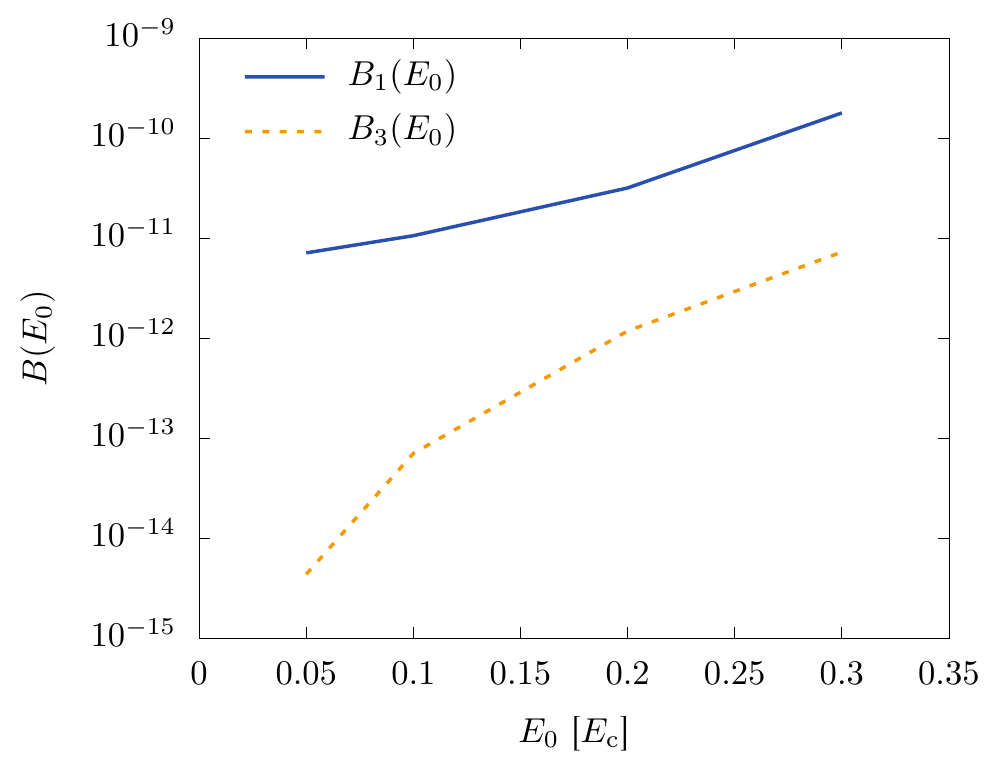}
\caption{Spectral energy density~(\ref{po}) of the plasma oscillations for $E_0=0.05E_\text{c}$, $0.1E_\text{c}$, $0.2E_\text{c}$, and $0.3E_\text{c}$ (left) and the relative brightness $B_1$ and $B_3$ defined in Eq.~(\ref{br}) as a function of $E_0$ according to this data (right). The other field parameters are $\Omega = 0.1 m$, $\sigma = 12$.
\label{fig:4}}
\end{figure*}
The currents $j^\text{cond}(t)$ and $j^\text{pol}_\text{R}(t)$ should now be governed by a combined action of the external and internal fields, i.e., by the effective field $E_\text{eff}(t)=E(t)+E_\text{int}(t)$. The backreaction problem requires a self-consistent description of the KE~(\ref{ke}) [or the KE system~(\ref{ode})] and Eq.~(\ref{Max}) (see Refs.~\cite{Kluger:1998, Bloch:1999} and references therein).

Here we are interested in the spectral energy density of the plasma oscillations,
\begin{equation}
\label{po}
w(\omega) = \frac{1}{8\pi}|E_\text{int}(\omega)|^2 = \frac{1}{8\pi\omega^2}|j(\omega)|^2
\end{equation}
that follows from Eq.~(\ref{Max}). 
The behavior of this function is shown in Fig.~\ref{fig:4}~(left) for four different values of the field amplitude $E_0$ ($\Omega = 0.1 m$, $\sigma = 12$).
The peak at the fundamental frequency of the external field and the third harmonic are well observed.
There are signs of odd harmonics of higher orders, but for quantitative estimates of their characteristics, the data available at our disposal are not sufficiently accurate.

To assess the response of the physical vacuum to the action of the external field, let us consider the ratio of the spectral energy density of the plasma field (\ref{po}) for the fundamental and third harmonics to the spectral energy density of the external field. Namely, we will explore here the relative brightness of the harmonics,
\begin{equation}
\label{br}
B_1(E_0) = \frac{|E_\text{int}(\Omega)|^2}{|E(\Omega)|^2},\quad B_3(E_0) = \frac{|E_\text{int}(3\Omega)|^2}{|E(\Omega)|^2},
\end{equation}
depending on the external field amplitude.

The results for the range $0.05 \le E_0/E_\text{c} \le 0.3$ are shown in the Fig.~\ref{fig:4}~(right). If the external field is present within a finite spatial region, the inner plasma field can escape from this domain, so one is able to detect the corresponding radiation far from the active area.
Of course, at the fundamental frequency of the external field, its energy absolutely dominates $|E_\text{int}(\Omega)|^2 \ll |E(\Omega)|^2$.
But the direction of propagation of the rays that form this field is strictly determined. Within the KE formalism, the problem of extracting the angular distribution of the radiation flux generated by the plasma field is very difficult and beyond the scope of this study. It can only be noted that this radiation will be concentrated in a plane perpendicular to the plane of the polarization of the primary rays since the external and internal electric fields are collinear. But there is no clear reason to believe that this secondary radiation will go strictly along the propagation line of the beams forming the focal spot (cf. Ref.~\cite{Gies:2018}). Therefore, the results presented indicate that approaching the critical field strength in the focal spot should lead to the scattering of a (very small) part $B_1$ of the energy of the primary beams at sufficiently large angles.
It can be noted that such scattering can be regarded as an analogue of Thompson scattering of photons by free charges. But in this case, the charge carriers are virtual electrons and positrons, the role of which becomes noticeable in near-critical fields. The presence of high-frequency harmonics in plasma radiation can be considered as one more firm indication of the nonlinear vacuum polarization effects in a strong field preceding EPP creation.

The absence of even harmonics in the spectra of the currents is a consequence of the maximal space-time symmetry of the physical vacuum as an active medium and is in accordance with the general theory of the response of nonlinear systems to external perturbations~\cite{Shen:2002}. We also point out that quantum photon emission due to absorption of an even number of external-field quanta is also prohibited according to Furry's theorem~\cite{Furry:1937}.

\section{Summary and Conclusion \label{sect:co}}

A characteristic feature of nonequilibrium EPP generated from  vacuum under the action of intense external fields is the appearance of strong electric currents and the corresponding inner-plasma electromagnetic fields (the back reaction effect~\cite{Bloch:1999,Kluger:1998}). Such phenomena can be described either in terms of the tadpole QED diagram~\cite{Karbstein:2015_2, Alex:2019, Gies:2018, Blinne:2019, Karbstein:2019, King:2018, DiPiazza:2005, Fedotov:PLA:2007, Narozhny:2007} or in the framework of a specific kinetic theory, as in the present study, predicting quasiclassical radiation. These two approaches are closely related: in the QED tadpole diagram the wave function of the outgoing photon is contracted with a loop which corresponds to the mean value of the current operator. Therefore, the spectral analysis of the current allows one to describe the process of photon emission. On the other hand, evaluating the mean value of the current operator in terms of the functions involved in the KEs~(\ref{ke}) and (\ref{ode}), one exactly obtains the expressions~(\ref{j_cond}) and (\ref{j_pol}). We also point out that this formalism does not incorporate quantum radiation due to annihilation of the $e^+e^-$ pairs in plasma (see Ref.~\cite{blaschke_prd_2011}). Another source of the EPP radiation is the process of photon emission accompanying production of $e^+ e^-$ pairs (see, e.g., Fig.~1 from Ref.~\cite{Alex:2019}).The first steps in studying this mechanism were taken in Refs.~\cite{DiPiazza:2005, Smol:2019, otto_prd_2017, Alex:2019}.

The present investigation was motivated, on the one hand, by the previous work of the authors (A.D.P. and S.A.S.)~\cite{Smol:2020} on the nonperturbative kinetic description of excitations in graphene under the action of an external time-dependent electric field and, on the other hand, by studies~\cite{Bowlan:2014, Yoshikawa:2017, Baudisch:2018}, where both theoretical and experimental investigations of the radiation processes in graphene in optical and infrared domains were carried out.
 
In this study, it was shown that under the action of a short linearly polarized laser pulse with a high energy density, plasma oscillations are excited in the physical vacuum. Their spectrum contains fundamental frequency of the external field and its high-frequency harmonics. It is expected that these vibrations cause additional radiation that escapes from the spatial region where the external laser fields are localized. The detection of this radiation and the observation of the third harmonic will convincingly indicate that the effects of vacuum polarization within strong-field QED take place in such a process, that is, it approaches the critical regime where the production of electron-positron pairs should become possible. Let us underline the fact that the results obtained are of a rather general and stable character with respect to the specific choice of the external field setting. For example, the same conclusion can be drawn in the case of a circularly polarized field as the physical vacuum can be regarded as a homogeneous isotropic active medium with a nonlinear response. Moreover, according to our data, the properties of the additional radiation are qualitatively independent of the Keldysh parameter $\gamma = m\Omega/ |eE_0|$, which obeys $\gamma \ll 1$ in the tunneling (Schwinger) regime and $\gamma \gg 1$ in the perturbative (multiphoton) regime, provided the external-field frequency $\Omega$ is well defined (in this paper, we presented the results for $\gamma = 0.3$--$2.0$).

\section*{Acknowledgments}
The work of A.D.P. and S.A.S. is supported by Russian Science Foundation (Grant No.~19-12-00042). I.A.A. acknowledges the support from the Russian Foundation for Basic Research (Grant No.~20-52-12017).

\end{document}